\documentclass[twocolumn,showpacs,preprintnumbers,amsmath,amssymb,superscriptaddress,PRL]{revtex4}
\usepackage{graphicx}
\usepackage{dcolumn}
\usepackage{bm}

\begin{document}
\topmargin -1.5cm

\title{Search for  $\Theta^{++}$ Pentaquarks  in the Exclusive Reaction $\gamma p\to K^+K^-p$}

\newcommand*{\RPI}{Rensselaer Polytechnic Institute, Troy, New York 12180-3590}
\affiliation{\RPI}
\newcommand*{\JLAB}{Thomas Jefferson National Accelerator Facility, Newport News, Virginia 23606}
\affiliation{\JLAB}
\newcommand*{\INFNGE}{Istituto Nazionale di Fisica Nucleare, Sezione di Genova, and Dipartimento di Fisica, Universit\'a di Genova,
 16146 Genova, Italy}
\affiliation{\INFNGE}
\newcommand*{\SCAROLINA}{University of South Carolina, Columbia, South Carolina 29208}
\affiliation{\SCAROLINA}
\newcommand*{\CMU}{Carnegie Mellon University, Pittsburgh, Pennsylvania 15213}
\affiliation{\CMU}

\newcommand*{\RICE}{Rice University, Houston, Texas 77005-1892}
\affiliation{\RICE}
\newcommand*{\UCONN}{University of Connecticut, Storrs, Connecticut 06269}
\affiliation{\UCONN}
\newcommand*{\ANL}{Physics Division, Argonne National Laboratory, Argonne, Illinois 60439-4843}
\affiliation{\ANL}
\newcommand*{\ASU}{Arizona State University, Tempe, Arizona 85287-1504}
\affiliation{\ASU}
\newcommand*{\UCLA}{University of California at Los Angeles, Los Angeles, California  90095-1547}
\affiliation{\UCLA}
\newcommand*{\CSU}{California State University, Dominguez Hills, California  90747-0005}
\affiliation{\CSU}
\newcommand*{\CUA}{Catholic University of America, Washington, D.C. 20064}
\affiliation{\CUA}
\newcommand*{\SACLAY}{CEA-Saclay, Service de Physique Nucl\'eaire, F91191 Gif-sur-Yvette, France}
\affiliation{\SACLAY}
\newcommand*{\CNU}{Christopher Newport University, Newport News, Virginia 23606}
\affiliation{\CNU}
\newcommand*{\ECOSSEE}{Edinburgh University, Edinburgh EH9 3JZ, United Kingdom}
\affiliation{\ECOSSEE}
\newcommand*{\FIU}{Florida International University, Miami, Florida 33199}
\affiliation{\FIU}
\newcommand*{\FSU}{Florida State University, Tallahassee, Florida 32306}
\affiliation{\FSU}
\newcommand*{\GWU}{The George Washington University, Washington, DC 20052}
\affiliation{\GWU}
\newcommand*{\ECOSSEG}{University of Glasgow, Glasgow G12 8QQ, United Kingdom}
\affiliation{\ECOSSEG}
\newcommand*{\ISU}{Idaho State University, Pocatello, Idaho 83209}
\affiliation{\ISU}
\newcommand*{\INFNFR}{INFN, Laboratori Nazionali di Frascati, 00044Frascati, Italy}
\affiliation{\INFNFR}
\newcommand*{\ORSAY}{Institut de Physique Nucleaire ORSAY, Orsay, France}
\affiliation{\ORSAY}
\newcommand*{\IHEP}{Institute for High Energy Physics, Protvino, 142281, Russia}
\affiliation{\IHEP}
\newcommand*{\ITEP}{Institute of Theoretical and Experimental Physics, Moscow, 117259, Russia}
\affiliation{\ITEP}
\newcommand*{\JMU}{James Madison University, Harrisonburg, Virginia 22807}
\affiliation{\JMU}
\newcommand*{\KHARKOV}{Kharkov Institute of Physics and Technology, Kharkov 61108, Ukraine}
\affiliation{\KHARKOV}
\newcommand*{\KYUNGPOOK}{Kyungpook National University, Daegu 702-701, Republic of Korea}
\affiliation{\KYUNGPOOK}
\newcommand*{\UMASS}{University of Massachusetts, Amherst, Massachusetts  01003}
\affiliation{\UMASS}
\newcommand*{\MOSCOW}{Moscow State University, General Nuclear Physics Institute, 119899 Moscow, Russia}
\affiliation{\MOSCOW}
\newcommand*{\UNH}{University of New Hampshire, Durham, New Hampshire 03824-3568}
\affiliation{\UNH}
\newcommand*{\NSU}{Norfolk State University, Norfolk, Virginia 23504}
\affiliation{\NSU}
\newcommand*{\OHIOU}{Ohio University, Athens, Ohio  45701}
\affiliation{\OHIOU}
\newcommand*{\ODU}{Old Dominion University, Norfolk, Virginia 23529}
\affiliation{\ODU}
\newcommand*{\URICH}{University of Richmond, Richmond, Virginia 23173}
\affiliation{\URICH}
\newcommand*{\UNIONC}{Union College, Schenectady, NY 12308}
\affiliation{\UNIONC}
\newcommand*{\VT}{Virginia Polytechnic Institute and State University, Blacksburg, Virginia   24061-0435}
\affiliation{\VT}
\newcommand*{\VIRGINIA}{University of Virginia, Charlottesville, Virginia 22901}
\affiliation{\VIRGINIA}
\newcommand*{\WM}{College of William and Mary, Williamsburg, Virginia 23187-8795}
\affiliation{\WM}
\newcommand*{\YEREVAN}{Yerevan Physics Institute, 375036 Yerevan, Armenia}
\affiliation{\YEREVAN}
\newcommand*{\UK}{University of Kentucky, Lexington, Kentucky 40506}
\affiliation{\UK}
\newcommand*{\UNCW}{University of North Carolina, Wilmington, North Carolina 28403}
\affiliation{\UNCW}
\newcommand*{\UAT}{North Carolina Agricultural and Technical State University, Greensboro, North Carolina 27455}
\affiliation{\UAT}
\newcommand*{\RIKEN}{The Institute of Physical and Chemical Research, RIKEN, Wako, Saitama 351-0198, Japan}
\affiliation{\RIKEN}
\newcommand*{\NOWUNH}{University of New Hampshire, Durham, New Hampshire 03824-3568}
\newcommand*{\NOWUMASS}{University of Massachusetts, Amherst, Massachusetts  01003}
\newcommand*{\NOWMIT}{Massachusetts Institute of Technology, Cambridge, Massachusetts  02139-4307}
\newcommand*{\NOWODU}{Old Dominion University, Norfolk, Virginia 23529}
\newcommand*{\NOWSCAROLINA}{University of South Carolina, Columbia, South Carolina 29208}
\newcommand*{\NOWGEISSEN}{Physikalisches Institut der Universit\"at Gie{\ss}en, 35392 Giessen, Germany}
\newcommand*{\NOWNONE}{unknown, }

\author {V.~Kubarovsky} 
\affiliation{\RPI}
\affiliation{\JLAB}
\author {M.~Battaglieri} 
\affiliation{\INFNGE}
\author {R.~De Vita} 
\affiliation{\INFNGE}
\author {J.~Goett} 
\affiliation{\RPI}
\author {L.~Guo} 
\affiliation{\JLAB}
\author {G.S.~Mutchler} 
\affiliation{\RICE}
\author {P.~Stoler} 
\affiliation{\RPI}
\author {D.P.~Weygand} 
\affiliation{\JLAB}
\author {P.~Ambrozewicz} 
\affiliation{\FIU}
\author {M.~Anghinolfi} 
\affiliation{\INFNGE}
% ...

\collaboration{The CLAS Collaboration}
     \noaffiliation

\date{\today}

\begin{abstract}
The reaction 
$\gamma p \rightarrow pK^+K^-$ 
was studied
at Jefferson Lab with photon energies from 1.8  to 3.8 GeV
using a tagged photon beam. The goal was to search for  a $\Theta^{++}$\ pentaquark,
a narrow doubly charged baryon state having strangeness S=+1  and isospin I=1, in the  
$pK^+$ invariant mass spectrum.
No statistically significant evidence of a $\Theta^{++}$  was found.
Upper limits on the total and differential production cross section for the reaction
$\gamma p \rightarrow K^-\Theta^{++}$ were  obtained   in the mass
range from 1.5 to 2.0 GeV/c$^2$, with an upper limit of about 0.15 nb, 95\% C.L. for a
narrow resonance with a mass $M_{\Theta^{++}}=1.54$ GeV/c$^2$. 
This result places a very stringent  upper limit  on the  $\Theta^{++}$ width.


\end{abstract}


\maketitle

Since the first reports of possible observations of a $\Theta^{+}$ pentaquark, 
a narrow baryon state having strangeness S=+1, there has been a
great deal of speculation about its isospin structure
\cite{walliser,capstick,binwu,ellis,roberts,kopeliovich,shi-lin,nishikawa,igor}.
If it were an isotriplet (I=1), one might expect to observe its isospin partners,
in particular $\Theta^{0}$ and $\Theta^{++}$.
For example, Roberts~\cite{roberts}  calculated the production of pentaquarks in 
an isobar  approach for all scenarios of spin, isospin and parity. It was
concluded that an isovector pentaquark should have a cross section comparable 
with an isoscalar pentaquark, 
and if the $\Theta^{+}$ is indeed isovector, one should expect to 
observe a comparable $\Theta^{++}$  cross section.  
However, on the experimental side, evidence  of the existence of a $\Theta^{++}$ had not 
been  forthcoming.
Gibbs ~\cite{gibbs},  analyzing available $K^+p$ total cross sections, finds no 
evidence for an isovector  resonance.
Meanwhile, the standard PWA of the elastic KN scattering
suggests \cite{hyslop}
at least two $\Theta^{++}$ states, but very broad and with
masses much higher than that of the $\Theta^+$.
Modification of this PWA \cite{igor} provides a whole set of
candidates for the $\Theta^{++}$, in particular, near the $\Theta^+$ mass.
But all those states, if confirmed, should be very narrow,
having elastic widths less than 0.1 MeV.

Other experiments  involving  electromagnetic probes,  
(CLAS~\cite{juengst,clas2},
ZEUS~\cite{zeus}, 
SAPHIR~\cite{saphir}, and  
HERMES~\cite{hermes}), 
reported  that 
no statistically 
significant signal for the $\Theta^{++}$  decaying to $pK^+$ was observed, even though each reported 
positive observations for  candidate  $\Theta^{+}$  peaks. 
Indeed, although positive observations  of the $\Theta^+$ were reported in several laboratories,
many others have seen no evidence of it, and its existence has become questionable.
It has been noted  that   most of the reported negative results were based 
on high energy experiments involving $pK^0$ invariant mass reconstructions within large 
backgrounds of reaction fragments. 
Thus, it was speculated that even if a pentaquark does exist, a possible 
interpretation for non-observation in such experiments is  that there would
be a small  probability for pentaquark  formation, either  $I$=0 or 1,  since
most of the quarks necessary to construct them would have to be created within 
the evolution of the reaction itself. However, recent reports of several  
exclusive follow-up experiments at lower
energies at CLAS~\cite{g11,g10}  failed to verify earlier reported  $\Theta^+$ pentaquark signals 
observed at  SAPHIR~\cite{saphir} and CLAS~\cite{g2} .
Moreover,  a $\Theta^{++}$ search at Jefferson Lab Hall-A~\cite{hanson} using
high-resolution limited-acceptance magnetic spectrometers studied the exclusive reaction 
$ep\to e^\prime K^-(\Theta^{++} )$ with no observation of a $\Theta^{++}$ peak.  
One can summarize that, unlike the situation of the $\Theta^{+}$,  which itself remains highly
controversial,  there had been no  positive reports of the possible existence of
a $\Theta^{++}$. 

Recently the issue of an isovector pentaquark was revitalized by the report of possible 
signals for  $\Theta^{++}$  and anti-$\Theta^{++}$ ($\bar \Theta^{++}$)   observed in d-Au and Au-Au 
interactions by the STAR Collaboration~\cite{star} at RHIC, in the invariant mass spectra
of detected $pK^+$ and $\bar p K^-$ pairs, respectively. The reaction was inclusive
since the high-energy beams (200 GeV/c for d-Au collisions) resulted in  very large final 
state multiplicities producing a large underlying combinatoric background. The subtraction
of the background   yielded
a 3.5$\sigma$ to 5$\sigma$ peak at 1.53 GeV/c$^2$.  Although the peak was observed in
d-Au collisions, there were no significant peaks observed from 
Au-Au and Cu-Cu collisions.
Also, in such a  highly inclusive reaction  it is impossible to know if there are any   
associated particles, and whether the production is in the direct
formation, or in the final state of a multi-step process.  Thus, it would
be advantageous to perform an experiment in which the
initial state is carefully controlled, the final state is exclusively measured, 
and with large acceptance to include as much phase space as possible. 

A  feature of all positive observations of pentaquarks is that the signals
contained very limited statistics. 
 In this letter we report the results of a high statistics search for the  production
 of the  $\Theta^{++}$  state in the reaction 
$\gamma p \to K^-\Theta^{++}$,  with
$\Theta^{++}\to pK^+$. 
 The experiment was performed at the Jefferson Lab CLAS facility. Details of the design and
 operation of the CLAS spectrometer and its components may be found in 
 Ref.~\cite{CLAS} and references within.  Reference \cite{PRD} discusses the experimental
 setup used in the present study in greater detail. 
 An energy  tagged bremsstrahlung  beam produced by 
a continuous 60~nA electron beam of energy $E_0=4.02$ GeV,
impinging on  
a gold radiator of thickness $8 \times 10^{-5}$ radiation lengths, yielded incident photons in the 
 energy range 1.8 to 3.8 GeV. The photon energy for each
 event was determined by  means of a tagger placed upstream of the CLAS spectrometer.
 The photon energy  resolution  was approximately  $0.1\% \times E_0$.
 The reaction target consisted of liquid hydrogen contained in a cylindrical mylar cell
 of length 40 cm. 
 
 Charged particles were detected by  the CLAS spectrometer.  Particle tracking 
 utilized multiwire drift chambers and a toroidal magnetic field.
 Particle identification was primarily obtained by comparing the particle momentum
 with that calculated from the track length
 and  flight time  between scintillator detectors around the target
 and scintillator detectors surrounding the CLAS spectrometer. The CLAS momentum 
 resolution is of the order of 0.5-1\% ($\sigma$) depending on
the kinematics. The detector's geometrical acceptance for positively charged particles
in the  relevant kinematic region is about 40\%, and  
several times  smaller  for 
low energy negative  hadrons, which can be lost at  forward angles because they are bent 
out of the acceptance by the toroidal field. For example, the number of  $\Lambda(1520)$
obtained by the reconstruction of  $pK^-$ events is almost an order of magnitude
smaller than the number reconstructed from the missing mass of  $K^+$ when the $K^-$ is not
required to be detected.
Coincidences between the photon tagger and two charged particles in the CLAS detector 
triggered the recording of the events. 
The interaction time between the incoming photon and the target was measured by the start counter
\cite{start_counter}, consisting of a set of 24 2.2 mm thick plastic scintillators surrounding 
the hydrogen cell.
An integrated luminosity of 70 pb$^{-1}$ 
was accumulated in  50 days of running.

\begin{figure}[h]
\includegraphics[width=3.3in, height=3.3in]{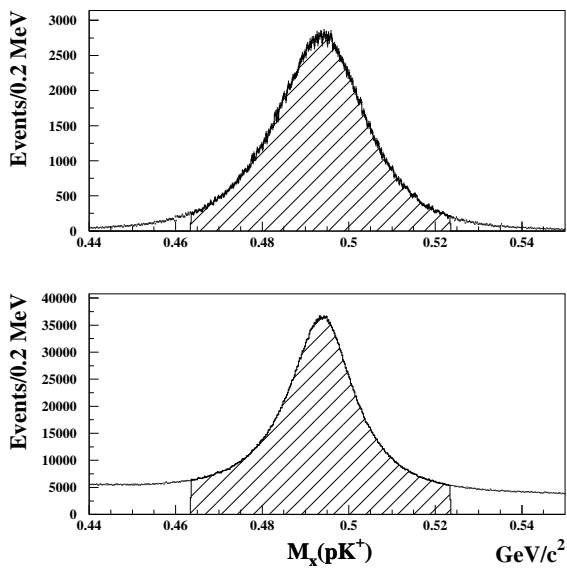}
\hspace{-0.3in}
\caption{The missing mass spectrum of $pK^+$ pairs. Upper panel: for events 
when all three final state particles, $p$, $K^+$ and $K^-$, are detected.
Lower panel: for events when only the $p$ and $K^+$ are required to be detected. 
Note the suppression
of background in the upper panel compared with the lower panel, in which the background 
is dominated by misidentification of pions as  kaons.
The shaded regions correspond to 3$\sigma$ cuts 
which define the region of retained events.  }
\label{mxpk}
\end{figure}

 Due to the high degree of exclusivity, the reaction $\gamma p \to K^-\Theta^{++}\to K^-K^+p$  
was studied in  two ways: 1) Requiring detection of  all three final state hadrons 
 $pK^-K^+$ and then directly observing the invariant mass of the $pK^+$.
 2) Detecting a $pK^+$ pair and identifying the $K^-$  by missing mass 
 reconstruction.  
 Fig.~\ref{mxpk}  compares the $K^-$ spectra obtained by each method.
The upper panel  shows the  missing mass spectrum of the detected $pK^+$
in Case 1, in which all three final state hadrons, $p$, $K^-$ and $K^+$, are required to be
detected. The complete dominance of the $K^-$ peak with  small background  indicates
that nearly all the events  are in the exclusive 3-body final state.
The lower panel  shows the  missing mass spectrum of the detected $pK^+$
in Case 2, in which only the $p$ and $K^+$ are required to be
detected. The significant background underlying the $K^-$ peak 
is mostly due to the misidentification of pions as kaons.
In both cases a cut of $3\sigma$ around the  $K^-$ peak was imposed on the
events which were retained for further analysis.

\begin{figure}
\includegraphics[width=3.5in]{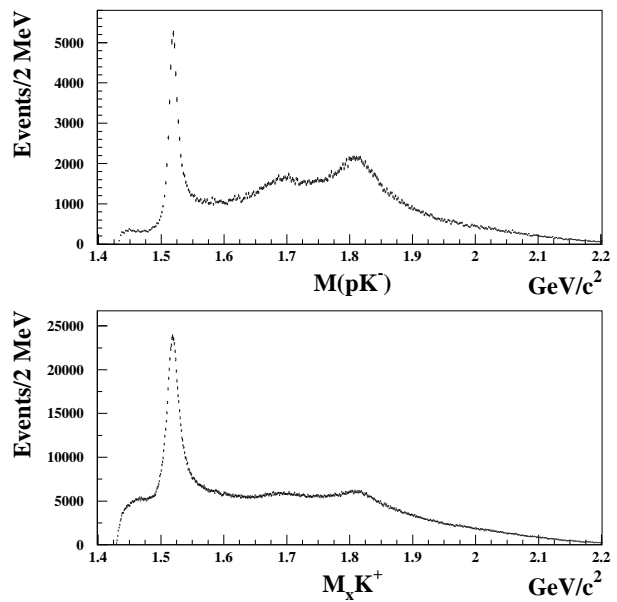}
\caption{Upper panel: The invariant mass spectrum
for the  $pK^-$  for events when all three final state particles, 
$p$, $K^+$ and $K^-$, are detected.
The most notable structure  is  the $\Lambda(1520)$ peak. Lower panel: 
The  $pK^-$ mass reconstructed
from the missing mass of the detected $K^+$ for events in which only the $p$ and 
$K^+$ are required to be detected. Events due to the $\Lambda(1520)$, as well as 
those due to  $\phi(1020)$ 
in the $K^+K^-$ spectra were removed from  further  analysis. The smaller number of 
$\Lambda(1520)$ events in the upper spectrum can be attributed to the much 
smaller acceptance for detection of the additional  $K^-$.    }
\label{mxk+}
\end{figure}

Figure~\ref{mxk+} shows the invariant mass of the $pK^-$ pairs after application of the cut on the $K^-$ peak.
The upper panel displays the invariant mass spectrum
for the  $pK^-$  for events for Case 1
and the lower panel displays the missing mass in Case 2. The most notable feature in each spectrum is the prominent  peak due to the 
$\Lambda(1520)$. In  comparing  the upper and lower panels of Figs.~{\ \ref{mxpk}} and~\ref{mxk+} 
it is observed that almost an order of magnitude in statistics is gained due to the increased acceptance
in requiring only the detection of 
the  $p$ and $K^+$.  There are nearly 1 million events corresponding to the $\Lambda(1520)$ peak 
for Case 2. The trade-off is that the pion contamination is significantly 
greater for Case 2 than Case 1. In addition  to  greater statistics, Case 2 has an advantage in that the  undetected
 $K^-$ can be emitted at any value of $\cos\theta_{CM}$, 
where   $\theta_{CM}$ is the angle between the electron beam
direction and $pK^+$ system in the center-of-mass system, so that the acceptance is significant
 in the entire range of $\cos\theta_{CM}$, from -1 to +1, and $t$-channel processes are not suppressed.
 On the other hand, in Case 1 the  acceptance in $\cos\theta_{CM}$ for detecting  the $K^-$s 
 becomes  smaller near $\cos\theta_{CM}$=+1, so that $t$-channel processes are suppressed.

In all further analysis, 3$\sigma$ cuts were applied in the $pK^-$ 
and $K^+K^-$ mass spectra to eliminate the contribution of the $\Lambda(1520)$ and $\phi(1020)$,
respectively.

\begin{figure}[h]

\includegraphics[width=3.5in]{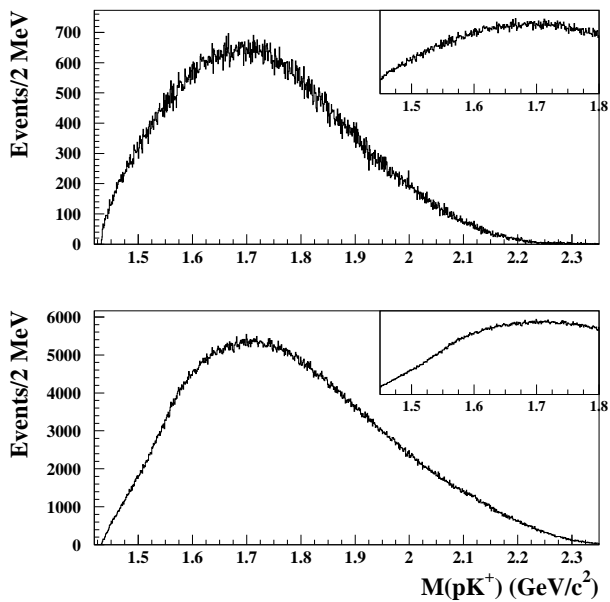}
\caption{The  $pK^+$  invariant mass  spectra
obtained after all cuts were applied,  including the removal of 
the $\Lambda(1520)$ and $\phi(1020)$ events.
Upper panel:  Case 1 in which
all three final state particles, $p$, $K^+$ and $K^-$, were detected.
Lower panel: Case 2, 
in which only the  $p$  and $K^+$ are required to be detected and the $K^-$ is identified
by missing mass cuts corresponding to the lower panel of Fig~\ref{mxpk}.  
The inset in each panel is a detail in the region near the reported $\Theta^+$ mass where one might 
expect a peak  due to the $\Theta^{++}$. 
In both cases  the spectra appear featureless. }
\label{mpk+}
\end{figure}

The $pK^+$ mass spectra  after all cuts were applied are shown for Case 1
and Case 2, in the upper and lower panels of Fig.~\ref{mpk+}, respectively.
In neither case is there any visual evidence for any narrow structures which could be
interpreted as due to a $\Theta^{++}$. The insets show expanded views
in the region where one might expect a $\Theta^{++}$ partner of an isovector  $\Theta^{+}$
located near $M=1.54$ GeV/$^2$. The  $pK^+$ mass resolution $\sigma(M_{\Theta^{++}})$ varies as 
a function of the mass from 
2 MeV/c$^2$  at $M_{\Theta^{++}}=1.5$~GeV/c$^2$, up to 
5.5 MeV/c$^2$ at $M_{\Theta^{++}}=2.0$~GeV/c$^2$. 
 
As for the $\Lambda(1520)$,  the acceptance of the undetected
$K^-$  is significant at all center-of-mass angles.
Thus, invariant mass 
spectra for $pK^+$ pairs were also obtained for discrete intervals of the center-of-mass 
angles of the emitted $K^-$ (or $pK^+$ pairs) covering the entire angular 
range.  No indication of a $\Theta^{++}$ peak is observed 
in any angular region.

Since no positive signal was observed,   upper limits for the cross sections 
were determined  for Case 2. Case 2 was chosen rather that Case 1 since
there are no gaps in the acceptance, and statistics is much higher.  Two  methods were
employed. In the first (Method 1), a Gaussian peak corresponding to   $N_ {\Theta^{++}}$,
and a polynomial background were fit  
to the $pK^+$  spectrum  for an assumed ${\Theta^{++}}$ mass, $M_{\Theta^{++}}$. 
Then a  Feldman-Cousins~\cite{feldman} algorithm was applied to the number under the fit 
peak and background in a $\pm 3\sigma$ interval  to obtain an upper limit of $\Theta^{++}$ events 
($N^{95\%}_ {\Theta^{++}}$) at the  95\% confidence level (CL) . 
This was repeated as a function of  $M_{\Theta^{++}}$.
In the second method (Method 2), the $pK^+$ spectrum was fit with a polynomial, excluding the region
of $M_{\Theta^{++}}$.  
For each $M_{\Theta^{++}}$  the $N_ {\Theta^{++}}$ was obtained as the difference between the polynomial 
and  the total number  of events
within a  $\pm3\sigma$   interval around $M_{\Theta^{++}}$.  Again, this
was analyzed with the Feldman-Cousins~\cite{feldman} algorithm. 
The cross section upper limit at the 95\% level was  then obtained from: 

$$
\sigma_{\Theta^{++}}^{95\%}={{N^{95\%}_{\Theta^{++}}}\over{L(M_{\Theta^{++}})\cdot \epsilon(M_{\Theta^{++}}) 
\cdot BR(\Theta^{++}\to pK^+)}},
$$

\noindent where $L(M_{\Theta^{++}})$ is the integrated luminosity for photons in the energy range from threshold
for a given mass  to 3.8 GeV,
$\epsilon(M_{\Theta^{++}})$ is the  $pK^+$ acceptance, and $BR(\Theta^{++}\to pK^+)$ is 
the branching ratio for 
$\Theta^{++}\to pK^+$, which is assumed to equal 1 for an isovector  $\Theta^{++}$.  

This procedure was repeated as a function 
of $M_{\Theta^{++}}$ and as a function of $\cos\theta_{CM}$ at  $M_{\Theta^{++}}=$1.54 GeV/c$^2$.
The upper limits obtained in Method 1 and Method 2 were found to be consistent.
Since the mass resolution $\sigma(M_{\Theta^{++}})$ varies approximately linearly, increasing with  
$M_{\Theta^{++}}$,
the variation in  $\sigma(M_{\Theta^{++}})$ and the acceptance $\epsilon(M_{\Theta^{++}})$ 
as a function of  $M_{\Theta^{++}}$  
were taken into account in determining the cross section upper limit $\sigma_{\Theta^{++}}^{95\%}$.
The CLAS acceptance,  $\epsilon(M_{\Theta^{++}})$ for the detection of the $\Theta^{++}$ 
was obtained by means of a detailed Monte Carlo simulation. The simulation assumed
$t$-channel dominance in which the $K^-$ is mainly produced at forward angles in the center-of-mass system. 
Assuming that the properties of the $t$-channel $K^-$ would be similar to that of the $K^+$
 in $\Lambda(1520)$ production,  
the energy dependence and the $t$-slope were taken from the experimental $\Lambda(1520)$ photoproduction
reaction. The Monte Carlo study showed that the acceptance was almost flat over the full
range of $\cos\theta_{CM}$. Thus, even for  extremely different  event generators, $t$-channel and 
$u$-channel $\Theta^{++}$  photoproduction,  the calculated acceptances differ by less than 10\%.
The result of the simulation is that the  CLAS acceptance with all the applied analysis cuts varied from 
6\%  at $M_{\Theta^{++}}=1.5$ GeV/c$^2$, up to
16\%  at $M_{\Theta^{++}}=2.0$ GeV/c$^2$.

The estimated systematic  errors in acceptance were combined with those of 
the detector inefficiencies, photon flux normalization and $\Theta^{++}$ mass resolution
to give an overall estimated 15\% systematic error in the resulting upper limit.
 
\begin{figure}[h]
\vspace{0.2in}
\includegraphics[width=3.0in,height=3.0in]{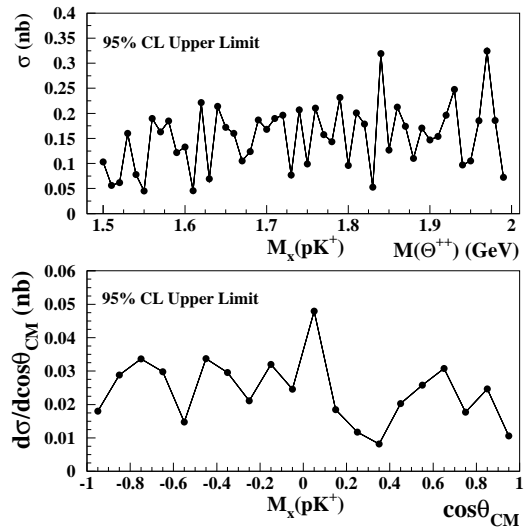}
\caption{Upper panel: The calculated upper limit on the cross section at 95\% 
confidence level vs. $M_{\Theta^{++}}$,
using the Feldman-Cousins approach, as discussed in the text,  
for Case 2 in which the $K^-$ was not required to be detected. The  upper limit 
at 95\% CL  at $M_{\Theta^{++}}$ near 1.54 GeV/c$^2$ is estimated to be approximately 0.15 nb.  
Lower panel: The upper limit as a function of $\cos\theta_{CM}$ at  $M_{\Theta^{++}}$ = 1.54 GeV/c$^2$. 
The systematic uncertainty in the magnitude of the upper limit is estimated at 15\%.}
\label{ul-FC}
\end{figure}

The resulting upper limit of the scans in  $M_{\Theta^{++}}$ and $\cos\theta_{CM}$
for Case 2 using Method 1 is shown in Fig.~\ref{ul-FC}. 
For both methods we find  the average upper limit in the mass region where a isospin 
partner of a $\Theta^+$ is expected,  near 1.54 GeV/c$^2$, at  approximately 0.15 nb,
and not much different   for the range of photon energies accessed in  this 
experiment  in the mass range from 1.5 to 2.0 GeV/c$^2$.

The upper limit of the ratio $\Theta^{++}/\Lambda(1520)$ was also obtained from the
data. The average cross section for $\Lambda(1520)$ photoproduction  was
calculated from the number of $\Lambda(1520)$ events in
 a manner similar to that described for  $\sigma_{\Theta^{++}}^{95\%}$ above. The result is 
$\sigma_{\Theta^{++}}/\sigma_{\Lambda(1520)}<2.3\times 10^{-4} $ at 95\% CL
averaged over the photon energy range of this experiment.

The $\Theta^{++}$ production cross section may be directly connected with the
$\Theta^{++}$ width $\sigma_{\Theta^{++}} \sim \Gamma_{\Theta^{++}}$ 
(see for example \cite{roberts}).
So small a  cross section implies a very narrow resonance width. However, an 
upper limit on the width would be highly model dependent, differing by as much 
as an order of magnitude for existing approaches~\cite{roberts,Ko,Oh,Nam,regge}.
For example, for an isovector pentaquark of $J^P=1/2^+$ the upper limit on the  width implied
by the present result for the Regge approach~\cite{regge} would be $\Gamma < 0.1$\  MeV/c$^2$, 
while for the effective Lagrangian approach~\cite{roberts} $\Gamma < 0.01$\  MeV/c$^2$.

In conclusion, the present experiment finds no evidence of the formation of a doubly charged
pentaquark in the exclusive channel $\gamma p \to K^-\Theta^{++} \to K^-K^+p$. The very high
statistical accuracy of the data allows one to obtain a rather small upper limit on
the cross section over a mass range from 1.5 to 2.0 GeV/c$^2$, with a  value of  about 0.15 
nb  at 95\% CL near 1.54 GeV/c$^2$  where a $\Theta^{++}$
isovector partner  of the $\Theta^{+}$ might be expected.  
Such small upper limits on the  cross section, and the implied width, makes it likely  that the
$\Theta^+$ baryon (if it exists)  has no isotopic partner and thus is an isosinglet state. 
Although the present experiment does put
very strong limits on the mechanisms which would be required to produce an isovector
pentaquark, we point out that it does not access a reaction in which a pentaquark
may be produced in association with an additional pion, as in Ref.~\cite{clas2}. 

We would like to thank 
W.~Roberts and  Ya.I.~Azimov  for valuable communications. 
We would like to acknowledge the outstanding efforts of the staff of the Accelerator
and the Physics Divisions at JLab that made this experiment possible. 
This work was supported in part by  the  Italian Istituto Nazionale di Fisica Nucleare, 
the French Centre National de la Recherche Scientifique and the 
Commissariat \`{a} l'Energie Atomique,  the U.S. Department of Energy, 
the National Science Foundation, 
and the Korea Research Foundation.
The Southeastern Universities Research Association (SURA) operates the
Thomas Jefferson National Accelerator Facility for the United States
Department of Energy under contract DE-AC05-84ER40150.


\begin{thebibliography}{99}
  

\bibitem{walliser}
H.~Walliser and V.~B.~Kopeliovich, ZhETF \textbf{124}, 483
(2003) [JETP, \textbf{97}, 433 (2003)].

\bibitem{capstick} S. Capstick, P. R. Page, and W. Roberts,
Phys. Lett.  {\bf B570}, 185 (2003).

\bibitem{binwu}
Bin Wu and Bo-Qiang Ma, Phys. Rev.  {\bf D69}, 077501 (2004).

\bibitem{ellis}
J.~Ellis, M.~Karliner, and  M.~Praszalowicz, JHEP
\textbf{0405}, 002 (2004).

\bibitem{roberts} 
W.~Roberts,
Phys. Rev.  {\bf C70}, 065201 (2004).
\bibitem{kopeliovich}
V.~B.~Kopeliovich, Uspekhi Fiz.\ Nauk \textbf{174}, 323 (2004)
[Physics-Uspekhi, \textbf{47}, 309 (2004)].

\bibitem{shi-lin}
Shi-Lin~Zhu, Phys.\ Rev.\ Lett.\ \textbf{91}, 232002 (2004).

\bibitem{nishikawa}
T.~Nishikawa {\it et al},
Phys. Rev.  {\bf D71}, 016001 (2005).

\bibitem{igor}Ya.~I.~Azimov {\it et al.}, Eur.\ Phys.\ J.\ \textbf{A26},
79 (2005). 

\bibitem{gibbs} W.R. Gibbs,  Phys. Rev.  {\bf C70}, 045208 (2004).

\bibitem{hyslop}J.S. Hyslop {\it et al.}, Phys. Rev.  {\bf D46}, 961 (1992).

\bibitem{juengst} 
 H.~G.~Juengst (CLAS Collaboration)
 Nucl.\ Phys.\  \textbf{A754}, 265c (2005).

\bibitem{clas2}V.~Kubarovsky {\it et al.}(CLAS Collaboration),
       Phys. Rev.  Lett. {\bf 92}, 032001 (2004).
       
\bibitem{zeus} S.~Chekanov {\it et al.} (ZEUS Collaboration),
Phys. Lett.  {\bf B591}, 7 (2004).

\bibitem{saphir} J. Barth {\it et al.} (SAPHIR Collaboration),   
Phys. Lett.  {\bf B572}, 127 (2003). 

\bibitem{hermes} A. Airapetian {\it et al.} (HERMES Collaboration), 
Phys.\ Lett.\  {\bf B585}, 213 (2004).

\bibitem{g11}M.~Battaglieri {\it et al.} (CLAS Collaboration), 
Phys. Rev. Lett {\bf 96}, 042001 (2006).

\bibitem{g10}B. McKinnon {\it et al.} (CLAS Collaboration),
arXiv:hep-ex/0603028. Submitted to Phys. Rev. Lett.


\bibitem{g2} S. Stepanyan {\it et al.} (CLAS Collaboration), 
Phys. Rev. Lett. {\bf 91}, 252001 (2003).

\bibitem{hanson}J.-O. Hansen (for Jlab Hall-A Collaboration), 
talk presented at the workshop  ``Pentaquark 2005", Jefferson Lab., Oct. 20-22, 2005.

\bibitem{star}
Huan Z. Huang (for the STAR Collaboration), arXiv:nucl-ex/0509037,
to appear in proceedings of International Conference on QCD
and Hadron Physics at Beijing, China, June 16-20, 2005.

\bibitem{CLAS}B.A.~Mecking {\it et al.} (CLAS Collaboration), Nucl. Inst. Meth.  {\bf A503}, 513 (2003).

\bibitem{PRD} R. De Vita  {\it et al.} (CLAS Collaboration),  in preparation.

\bibitem{start_counter} Y.G. Sharabian {\it et al.},
Nucl. Inst. Meth.   {\bf A432}, 265 (2006).

\bibitem{feldman} G.J. Feldman and R.D. Cousins, Phys. Rev. {\bf D57}, 3873 (1998).

\bibitem{regge} H. Kwee {\it et al.}, Phys. Rev.  {\bf D72}, 054012 (2005).

\bibitem{Ko} C.M. Ko and W. Liu, arXiv:nucl-th/0410068.

\bibitem{Oh} Y. Oh {\it et al.}, Phys. Rept. {\bf 423}, 49-89 (2006).

\bibitem{Nam} S. Nam  {\it et al.}, Phys. Lett.  {\bf B633}, 483-487 (2006).

\end{thebibliography}
\end{document}